\documentclass[aps,prl,twocolumn,showpacs,floatfix,superscriptaddress]{revtex4}
\usepackage[]{graphicx}
\usepackage{epstopdf}

\usepackage{graphicx}
\usepackage{dcolumn}
\usepackage{bm}
\usepackage{graphicx}
\usepackage{xcolor}
\usepackage{amsmath}
\usepackage{amssymb}

\usepackage{siunitx}
\usepackage{enumerate}

\begin{document}

\title{
Quantum enhanced non-interferometric quantitative phase imaging}

\author{Giuseppe Ortolano}

\affiliation{Quantum metrology and nano technologies division,  INRiM,  Strada delle Cacce 91, 10153 Torino, Italy}
\affiliation{DISAT, Politecnico di Torino, Corso Duca degli Abruzzi 24,
10129 Torino, Italy}

\author{Alberto Paniate}
\affiliation{Quantum metrology and nano technologies division,  INRiM,  Strada delle Cacce 91, 10153 Torino, Italy}
\affiliation{DISAT, Politecnico di Torino, Corso Duca degli Abruzzi 24,
	10129 Torino, Italy}

\author{Pauline Boucher}
\author{Carmine Napoli}
\affiliation{Quantum metrology and nano technologies division,  INRiM,  Strada delle Cacce 91, 10153 Torino, Italy}

\author{Sarika Soman}
\author{Silvania F. Pereira}
\affiliation{ Imaging Physics Dept. Optics Research Group, Faculty of Applied Sciences, Delft University of Technology, Lorentzweg 1, 2628CJ Delft, The Netherlands}

\author{Ivano Ruo Berchera}
\affiliation{Quantum metrology and nano technologies division,  INRiM,  Strada delle Cacce 91, 10153 Torino, Italy}

\author{Marco Genovese}
\affiliation{Quantum metrology and nano technologies division,  INRiM,  Strada delle Cacce 91, 10153 Torino, Italy}

\begin{abstract}

Quantum entanglement and squeezing have significantly improved phase estimation and imaging in interferometric settings beyond the classical limits. However, for a wide class of non-
interferometric phase imaging/retrieval methods vastly used in the classical domain e.g., ptychography and diffractive imaging, a demonstration of quantum advantage is still missing. Here, we fill this gap by exploiting entanglement to enhance imaging of a pure phase object in a non-interferometric setting, only measuring the phase effect on the free-propagating field. This method, based on the so-called “transport of intensity equation", is quantitative since it provides the absolute value of the phase without prior knowledge of the object and operates in wide-field mode, so it does not need time-consuming raster scanning. Moreover, it does not require spatial and temporal coherence of the incident light. Besides a general improvement of the image quality at a fixed number of photons irradiated through the object, resulting in better discrimination of small details,  we demonstrate a clear reduction of the uncertainty in the quantitative phase estimation. 

Although we provide an experimental demonstration of a specific scheme in the visible spectrum, this research also paves the way for applications at different wavelengths, e.g., X-ray imaging, where reducing the photon dose is of utmost importance.

\end{abstract}	

\maketitle

\section{Introduction} 

Quantum imaging \cite{Berchera_2019, Moreau2019, Genovese_2016} and sensing \cite{Degen_2017, Pirandola_2018, Petrini_2020} have provided genuine and valuable advantages in many measurement applications ranging from fundamental physics \cite{Aasi_2013,Pradyumna_2020} to biology \cite{Taylor_2016,Casacio_2021, Petrini_2022} from microscopy  \cite{Schwartz_2012, Monticone_2014,Samantaray_2017, Tenne_2019}to optical sensors \cite{Lawrie2019,Lee2021}.

In particular, given the importance of optical phase measurement, appearing in all the science fields, a considerable effort has been made to exploit quantum entanglement or squeezing for this task. Quantum phase estimation through first-order interference involving the mixing of two optical modes in a linear \cite{Polino_2020, Demkowicz_2014}  or non-linear  \cite{Chekhova_2016, Kalashnikov_2016,Topfer_2022} interaction is well understood. The ultimate uncertainty bound with quantum optical states is known to scale with the number of probing particles $N$ as  $N^{-1}$, the so-called 'Heisenberg scaling'. In contrast, for the classical probing state, it is limited to  $N^{-\frac{1}{2}}$, referred to as the standard quantum limit (SQL) or shot noise limit. Although the quantum advantage would, in principle, be disruptive for  $N\gg1$ in a realistic scenario, the gain over the SQL is rather, in the form of a constant depending on the optical losses \cite{Demkowicz_2014}. Proofs of principle of quantum-enhanced linear interferometry with the so-called entangled NOON states have been achieved, for example, in phase contrast \cite{Ono_2013} and polarization scanning microscopy \cite{Israel_2014}, usually limited to the case of N=2. However, the generation and preservation of NOON states involving a higher number of particles are extremely demanding, so their usability in a real-world application is questionable. More practical is the use of squeezed states \cite{Caves_1963, Xiao_1987, Aasi_2013, Schnabel_2017, Gatto_2022}. Non-linear interferometry involving a parametric amplifier instead of a beam splitter for mixing the light mode is promising for some applications, especially because the detection can be done at a wavelength different from the probing one \cite{Kalashnikov_2016,Topfer_2022} and the quantum advantage is independent of the detection noise \cite{Manceau_2017}.
Moreover, with some remarkable exceptions \cite{Camphausen_2021, Frascella_2019}, these interferometric schemes do not provide spatial resolution or require raster scanning for extended samples. 

Other phase imaging methods born in the quantum domain exploit second-order intensity correlation (or two-photon coincidence) among signal and idler beams of SPDC to retrieve the phase information. In contrast, the first-order intensity measurement of either the signal or the idler arm does not show interference \cite{Gatti_2004}. These techniques include ghost imaging and diffraction \cite{Strekalov_1995, Valencia_2005, Zhang_2005, Shapiro_2012, Losero_2019}  quantum holography \cite{Lemos_2014, Vinu_2020,Devaux_2019,Defienne2021}, quantum Fourier ptychography \cite{Aidukas_2019} and phase reconfigurable contrast microscopy \cite{Hodgson_2023}. In general, the signal-to-noise ratio (SNR) is smaller compared to direct first-order measurement if considering the same number of impinging photons on the object. However, some advantages can be found in some cases at few photon illumination levels, for example, the rejection of independent external noise \cite{Meda_2017,Erkmen_2009, Brida_2011, Morris_2015}, robustness through turbulence and scattering \cite{Dixon_2011,Bina_2013}.  

Here we present a quantitative non-interferometric quantum-enhanced phase imaging (NIQPI) scheme exploiting quantum correlations that do not belong to any of the techniques mentioned above since it does not involve either interference or second-order intensity correlations. In fact, only first-order intensity in both branches is measured, so the full-field phase retrieval is obtained in real-time by single-shot measurement. We will demonstrate, both theoretically and experimentally that the method can provide a clear advantage compared to the corresponding classical direct imaging thanks to the quantum correlation. 

\begin{figure}[th]
	\centering
	\includegraphics[width=0.5\textwidth]{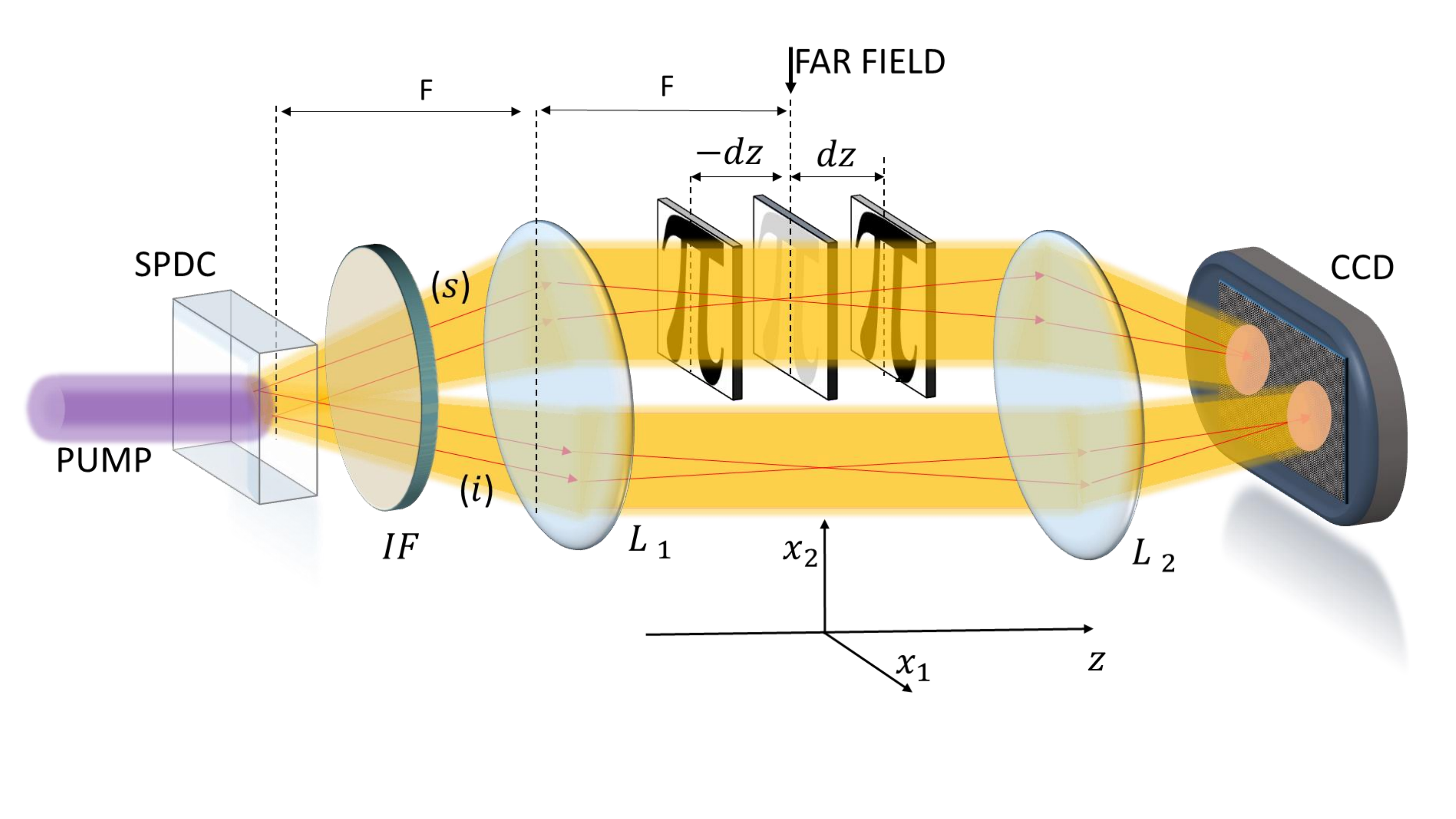}
	\caption{\textit{Scheme of the NIQPR}.  Two correlated beams labeled signal ($s$) and idler ($i$) are generated by the spontaneous parametric down conversion (SPDC) pumped by a CW laser @405nm and propagate through an imaging system composed of two lenses ($L_{1}$ is the far field lens with focal length $F=1$ cm and $L_{2}$ is the imaging lens with focal length of 3 cm) and a test object. An interference filter (IF) is used to select a bandwidth of 40 nm around the degenerate wavelength (@810nm) and to block the pump. $L_{2}$ images the far field plane on the camera chip around the focal plane with a magnification factor of about 8. The object is placed near to the far field of the source, and only the probe beam interacts with it. Phase information can be retrieved from intensity measurements  taken at some out of focus ($\pm dz$) planes.}\label{scheme}
\end{figure}

The NIQPI protocol exploits the scheme depicted in Fig. \ref{scheme}. We consider two quantum correlated beams produced by the spontaneous down-conversion process (SPDC), usually dubbed as signal beam ($s$) and idler ($i$) beam,  with intensity patterns that are perfectly identical in the far-field, point-by-point. Even the shot noise fluctuation is, in principle, perfectly reproduced in the two beams, which is impossible in the classical domain. The far field of SPDC is imaged at the sensors of a highly efficient and low-noise CCD camera. Only the signal beam probes the object, while the idler one is used as the reference for the noise. When the object is placed close to the far field but not exactly there, it produces an intensity perturbation on the signal photons propagation that is registered at the CCD camera. In particular, by measuring the signal intensity pattern $I(\bm{x},\pm dz)$ at the detection plane for two different 'defocused' object positions along the $z$-axis, namely, $+dz$ and $-dz$,  it is possible to reconstruct the phase profile  $\phi(\bm{x},z=0)$, by solving the so-called transport of intensity equation (TIE) \cite{Teague_1983}:
\begin{equation}
-k\frac{\partial}{\partial z}I(\bm{x},z)=\nabla_{\bm{x}}\cdot\left[I(\bm{x},0)\nabla \phi(\bm{x},0)\right]
\label{tie}
\end{equation}
where the derivative is approximated by the finite difference of two measurements out of focus, $\frac{\partial}{\partial z}I(\bm{x},z)\approx [I(\bm{x},d z)-I(\bm{x},-d z)]/(2 d z)$ and $I(\bm{x},z=0)$ is the far field intensity of the source. 
TIE is experimentally easy and computationally efficient as compared to the conventional phase retrieval techniques and, under suitable hypotheses, the method leads to a unique and quantitative wide-field image of the phase profile \cite{Teague_1983, Paganin_1998, Zuo_2020}. However, the reconstruction obtained in the signal arm can be strongly affected by the detection noise and by the shot noise if low illumination is used (see $M\&M$ for a detailed discussion). On the one hand, a faithful reconstruction through Eq. (\ref{tie}) requires a small defocus distance $|dz|$ in order to well approximate the derivative on its right-hand side. But, on the other hand, if $|dz|$ is too small, the effect of the phase gradient on the measured intensity becomes negligible and can be covered entirely by the shot noise. Here, we show that the noise pattern acquired on the idler beam can be used to reduce the effect of the shot noise in the signal beam, enhancing the overall phase image reconstruction and reducing the uncertainty on the quantitative spatially resolved phase retrieval \cite{Ortolano_2019}.
NIQPI can work with partially coherent light and has some advantages compared to interferometric schemes: it can be directly applied to wide-field transmission microscopy settings and it is intrinsically more stable than an interferometric setup \cite{Zuo_2020}. Moreover, since NIQPI is based on free propagation effect, it can be realized without using lenses and optical components, thus being particularly suitable in extreme UV or X-Ray imaging, where optical components are not efficient but where SPDC sources are available  and quantum enhanced detection has been already demonstrated \cite{Borodin2016,Sofer_2019}.

\begin{figure}[th]
	\centering
	\includegraphics[width=0.5\textwidth]{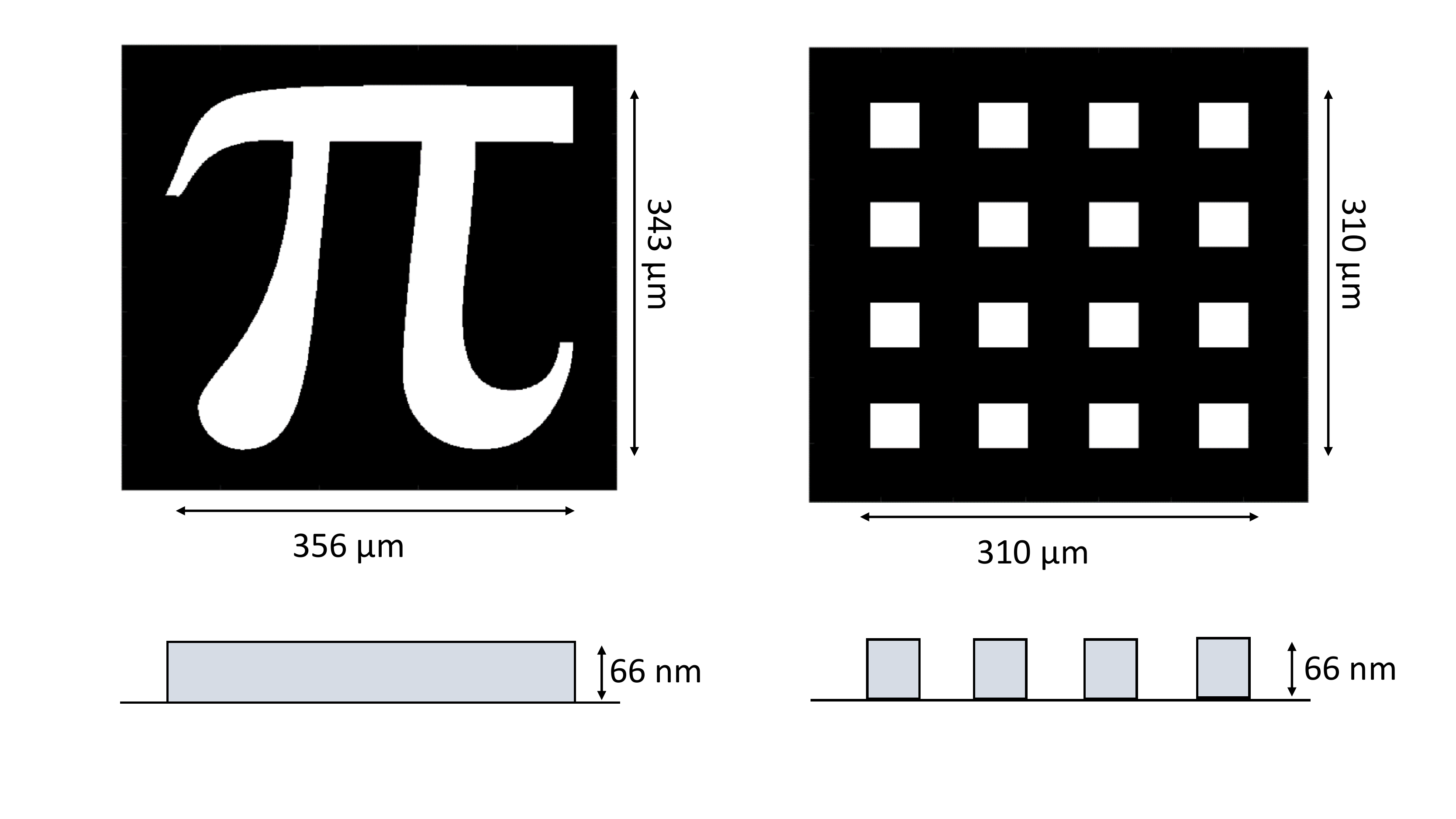}
	\caption{\textit{Sample}. Pure phase objects used in the experiment are sketched.  }\label{sample}
\end{figure}
\begin{figure*}[t]
	\centering
	\includegraphics[width=\textwidth]{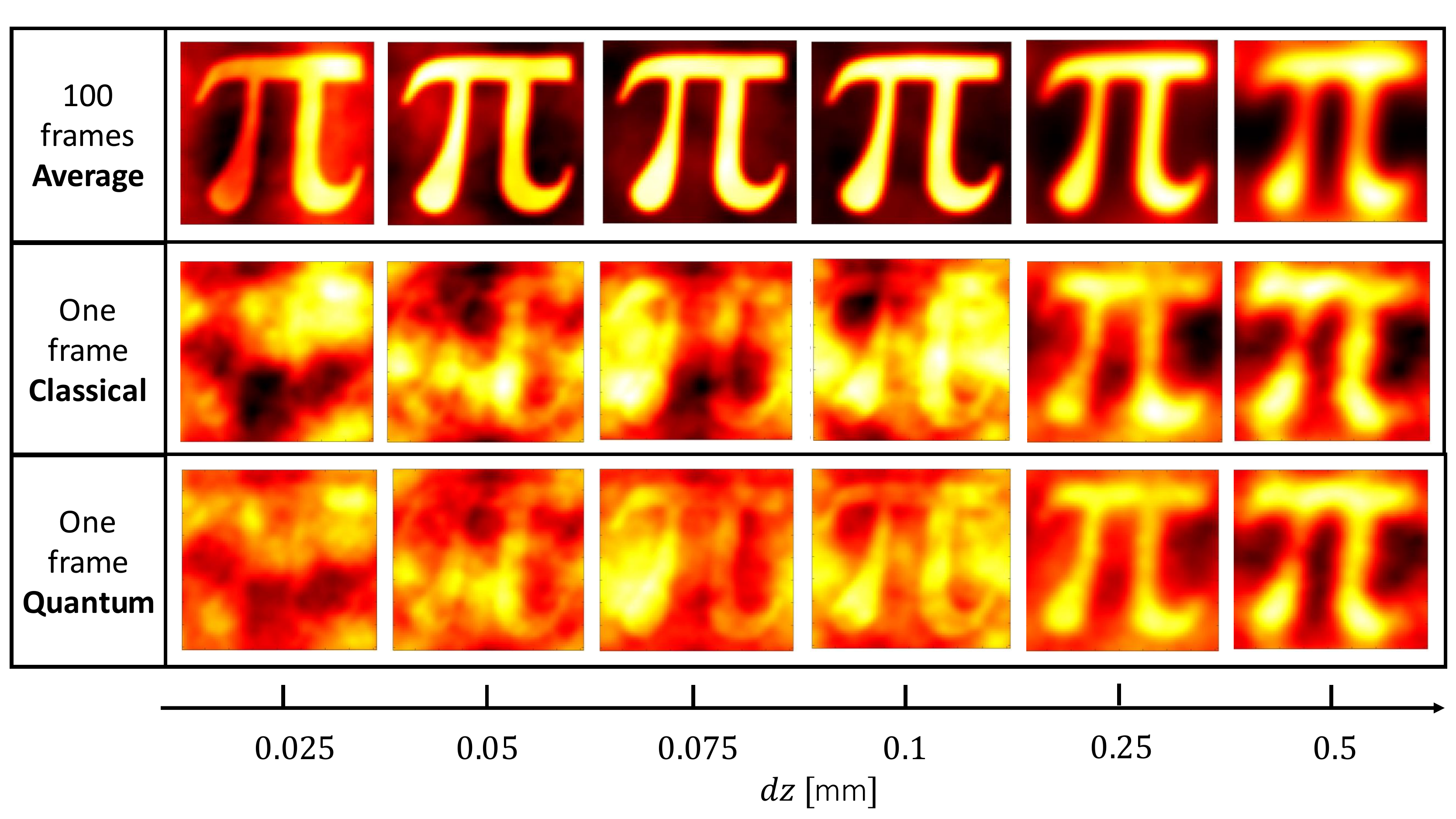}
	\caption{\textit{Experimental Reconstruction of the $``\pi$" sample as a function of the defocusing distance}. First row presents the phase reconstruction when 100 intensity patterns are used. Second and third rows show the single frame reconstructions for the classical and the quantum case, respectively. The size of each image is $80\times80$ $\text{pix}^{2}$}\label{pi}
\end{figure*}
\section{Results}

In our experiment, the number of photons per pixel per frame is about $n\approx10^3$, so that for the purpose of this work we can substitute the continuous quantity $I(\bm{x})$ appearing in Eq. (\ref{tie}) with the number of photons detected by the pixel at the coordinate $\bm{x}$.  Actually,  before the TIE algorithm, we apply an averaging filter of size $d=4$ to the intensity image, that consists in replacing the count in each pixel by the average count of a neighborhood of size $4\times4 \;\text{pix}^{2}$ around it, so that the final image conserves the same number of pixels. However, the averaging filter does not have any influence on the classical reconstructions, neither positive nor negative, while it improves the quantum reconstruction (see discussion in M$\&$M and related Fig. \ref{Pearson_vs_d}). From now on we will refer to $I(\bm{x})$ with that meaning, namely after the application of such averaging filter.

It is essential to point out that the SPDC source operates in the regime of very low photon number per spatio-temporal mode. In this limit, the photon statistics follows a Poisson distribution (see M$\&$M Sec.\ref{Experimental details: source, sample, detection} for details). So, aside from the negligible contribution of electronic readout noise, the measurement on the single beam is shot noise limited. 

We image pure phase objects reported in Fig. \ref{sample} with 66 $\pm$ 3 nm thickness estimated by profilometer (DektakXT, Bruker). It corresponds to a phase shift of 0.230 $\pm$ 0.012 rad @ 810 nm, the central degeneracy frequency of the SPDC photons. The samples have been realized by etching structures on a UV-fused Silica glass window using buffered oxide etch. 

Fig. \ref{pi} shows the experimental reconstructions of the $``\pi$"-shaped phase sample of Fig. \ref{sample} as a function of the defocussing distance $dz$. Each pixel of the phase image corresponds to a transverse resolution of about $5\mu$m in the object plane. As a reference, the first row of Fig. \ref{pi} shows the phase retrieved averaging 100 shots, so the shot noise effect is estimated to be negligible compared to the other sources of disturbance. However, even in this case, the reconstruction at small $d z$ is not perfect because of the well-known sampling error due to the discretization of the image, while at large defocussing the finite approximation of the derivative in $z$ fails, essentially producing blurred images. These two opposite trends determine a defocussing distance for which the reconstruction is optimal. The second row of Fig. \ref{pi} shows the reconstructions obtained by single frame intensities $I_{s}(\bm{x_{s}},\pm d z)$ measured at the CCD camera in the signal arm. In this case, the shot noise dominates and yields a drop in the reconstruction quality for all values of $d z$. How the noise on the intensity propagates to the phase reconstruction through the TIE is discussed in M\&M. In particular, the region of smaller $d z$ is the most affected since the intensity variation produced by the phase gradient is still small and is almost completely hidden in the shot noise. 

In order to take advantage of the quantum correlations here, we propose to replace into the Eq. (\ref{tie}) the single beam intensity with the following one \cite{Moreau_2017,Losero_2018,RuoBerchera_2020}:

\begin{equation}\label{I(s-i)}
I_{s-i}(\bm{x}, z)= I_{s}(\bm{x_{s}},z)- k_{opt} \delta  I_{i}(\bm{x_{i}},0).
\end{equation}
where $\delta  X\equiv \langle X\rangle-X$ represents the quantum fluctuation of the operator $X$, and $\langle\cdot\rangle$ is the quantum expectation value. In fact, the second term in Eq. (\ref{I(s-i)}) is meant to compensate for the quantum fluctuation of the signal pattern exploiting the local correlation between probe and reference beams. The factor $k_{opt}$ is a parameter chosen to minimize the residual fluctuation $\langle \delta^2 I_{s-i}\rangle$ and can be evaluated experimentally by calibration of the system since it is related to the detection efficiency. A phenomenological model describing noise reduction is discussed in M$\&$M. It turns out that the fluctuation of the quantity in Eq. (\ref{I(s-i)}) is reduced with respect to the shot noise according to the following expression: 

\begin{eqnarray}
\langle \delta^2 I_{s-i}\rangle&=& \left[1-\left( 1-\alpha\right)^{2}\eta ^2\right]\langle I(0)\rangle, \label{noisered}
\end{eqnarray}
where $0<\eta<1$ is the heralding efficiency, namely the probability of detecting an idler photon in the pixel in $\bm{x}_{i}$ conditioned to the detection of the correlated signal photon in the pixel in $\bm{x}_{s}$ (see M$\&$M section). The parameter $\alpha$ is the average fraction of photons that deviate from the original path due to the phase object and depends on the average phase gradient. It can be experimentally evaluated as the spatial average of the quantity $\alpha(\bm{x})\equiv\langle |I(\bm{x}_{s},0)-I(\bm{x}_{s},dz)|\rangle/\langle I(\bm{x}_{s},0)\rangle$. Eq. (\ref{noisered}) states that the intensity  fluctuation is reduced below the shot noise by a factor that depends on the efficiency in detecting quantum correlation and that it is effective if the object is weakly affecting the intensity distribution, namely when $\alpha\ll1$. In our experiment, following the absolute calibration method reported in \cite{Meda_2015, Avella_2016, Samantaray_2017}, we estimate $\eta=0.57$ for the particular case of averaging filter size $d=4$. The value of $\alpha$ for the faint object considered is very small, for example we estimated $\alpha=7 \cdot 10^{-3}$ for $dz=0.1$ mm.

The third row of Fig. \ref{pi} reports the reconstructions when the shot noise has been reduced using quantum correlations between probe and reference, according to Eq. ($\ref{noisered}$).  A general  improvement of the reconstruction can be appreciated. As expected, the noise reduction is more evident at smaller $dz$ leading to an improvement in the reconstruction of higher spatial frequency.  

\begin{figure}[th]
	\centering
	\includegraphics[height=0.38\textwidth, width=0.5\textwidth]{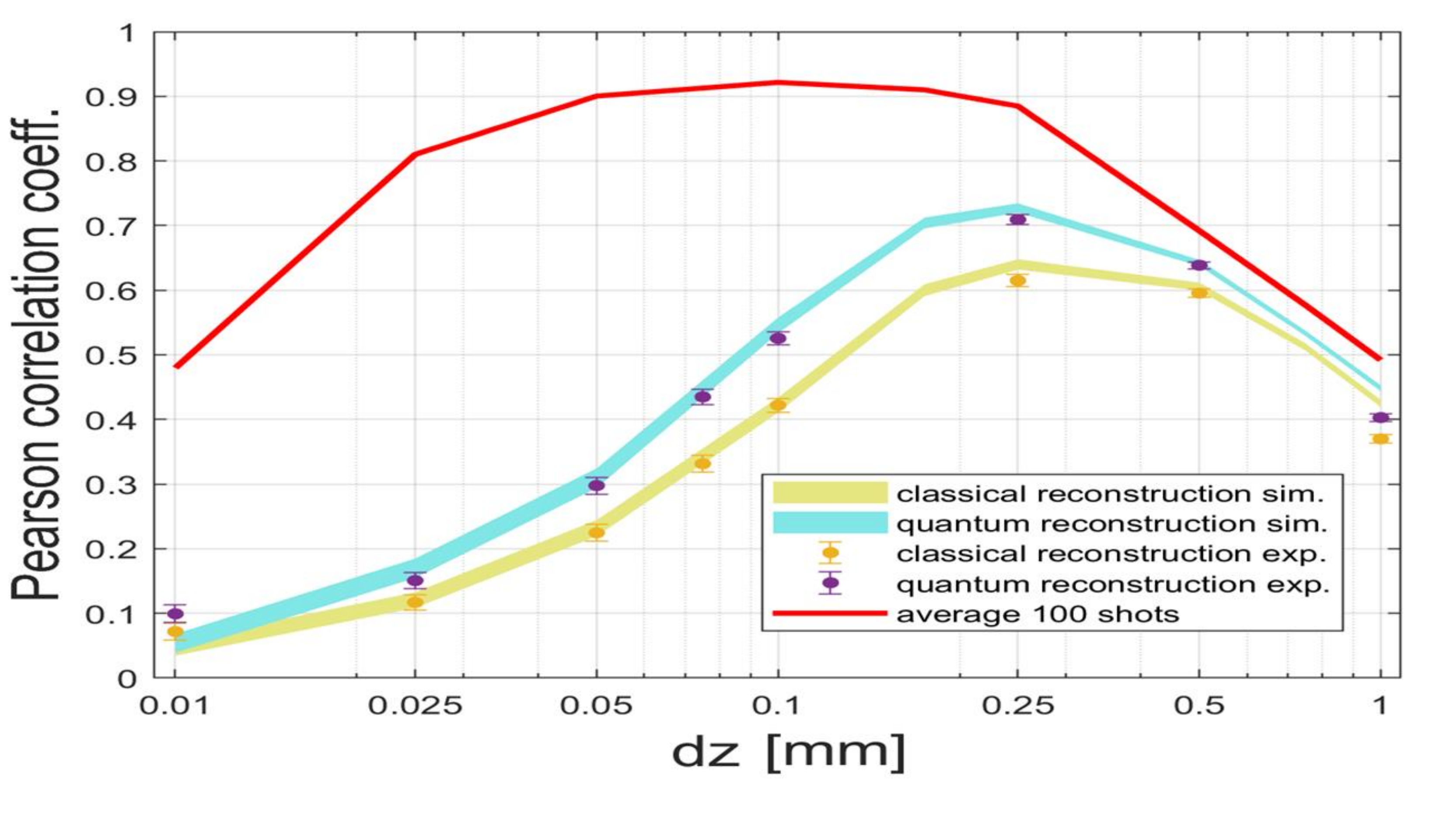}
	\caption{\textit{Pearson correlation between reconstructed and reference images}. The light-blue and yellow curves are the result of a Fourier optics based simulation. The line-width is the confidence interval of one standard deviation after an average over 100 reconstructions for each $dz$. The experimental points are represented as purple and yellow dots with uncertainty bar also corresponding to one standard deviation. The red curve corresponds to the reconstruction obtained by summing of 100 intensity patterns, where the shot noise becomes negligible (in this case, quantum and classical correlation overlaps). }\label{Pearson_vs_dz}
\end{figure}

A quantitative analysis of the quality of the reconstructions and of the quantum advantage can be performed by evaluating the Pearson correlation coefficient between the reference phase image and the reconstructed one. The Pearson coefficient is defined as,

\begin{equation}
\mathcal{C}= \frac{\sum_{\bm{x}}(\phi_{r}(\bm{x})-\bar{\phi_{r}})(\phi(\bm{x})-\bar{\phi})}{\sqrt{\text{Var}[\phi_{r}]\text{Var}[\phi]}}
\end{equation}
where $\bar{\phi}$ and Var$[\phi]$ denote the spatial mean and variance of the phase image $\phi$, and provides a simple and commonly used figure of merit to quantify the similarity between the two images. 
Fig. \ref{Pearson_vs_dz}, shows the Pearson coefficient as a function of the defocusing. Each curve has a correspondence with each image strip in Fig. \ref{pi}. The red curve corresponds to the reconstruction using 100 frames, where shot noise is negligible (corresponding to the first strip in Fig. \ref{pi}). The lower curves present the performance of single frame experimental reconstructions, both quantum and classical, obtained from a simulation. Experimental points are well in agreement with these simulations. As expected, according to this figure of merit, an optimal reconstruction is reached for the intermediate value of defocusing. The quantum advantage is confirmed in terms of correlation with the reference image.
 
 \begin{figure*}[th]
 	\centering
 	\includegraphics[height=0.38\textwidth, width=\textwidth]{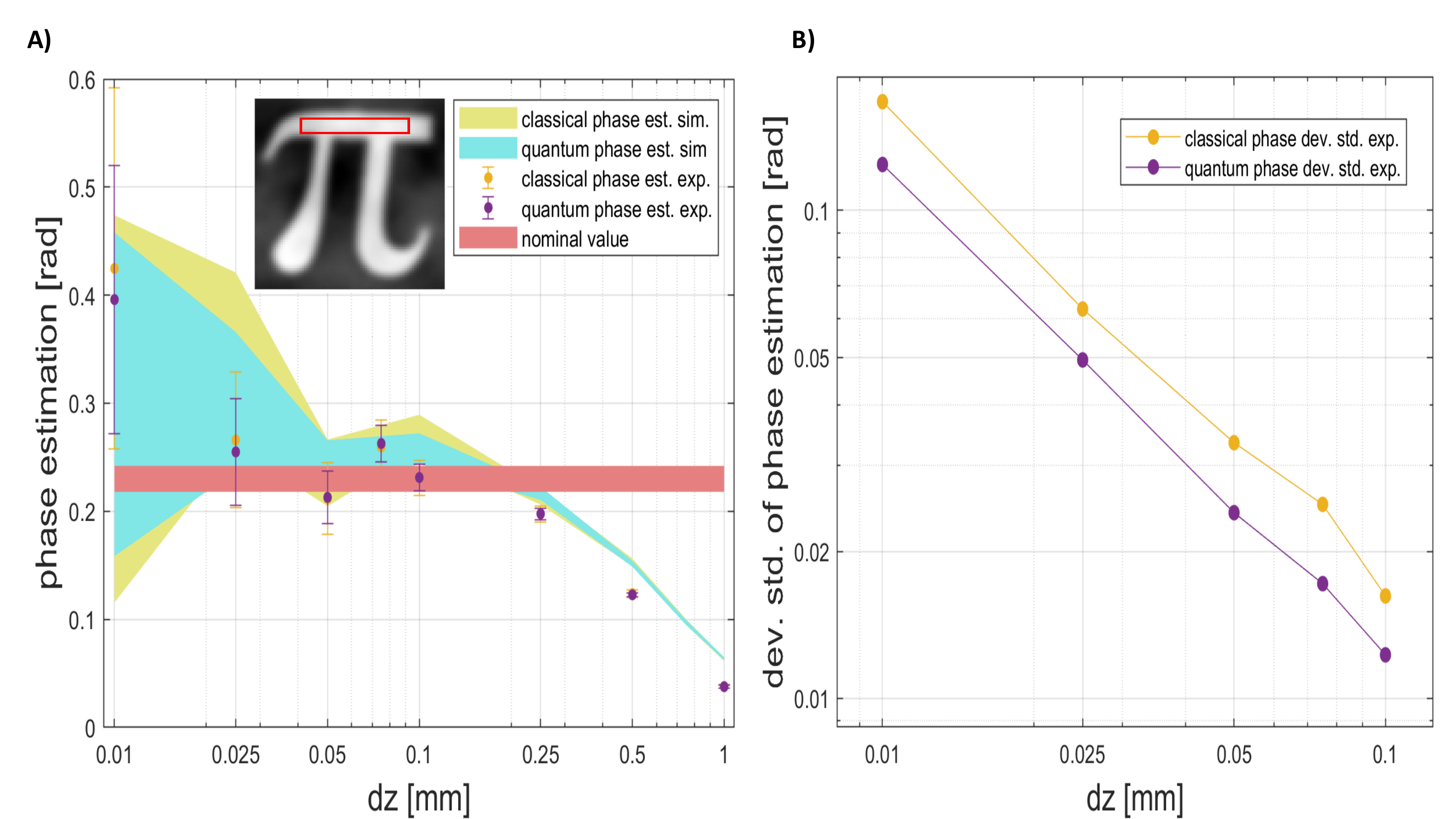}
 	\caption{\textit{Phase estimation}. \textbf{A} The estimated value of the phase step (average of the rectangular selected region) is plotted at different defocusing distances. Experimental points for the classical (yellow dot) and the quantum (purple dot) phase retrieval are compared with the simulations (reported for one standard deviation confidence band). For comparison, we also report the nominal value, estimated by the profilometer in reflection, of the phase step difference between the etched/non-etched areas.  \textbf{B} The uncertainty in the phase estimation for quantum and classical cases demonstrate the quantum advantage. }\label{Phase_estim}
 \end{figure*}
 
Besides the correct reconstruction of the complex phase profile assessed by the correlation coefficient, in many cases, it is of utmost importance to achieve a quantitative estimation of the phase. Fig. \ref{Phase_estim}A reports the phase value estimated as a function of $dz$, where, for the analysis, we have selected the region indicated in the red rectangle in the insets. The results indicate that the phase step is reconstructed without bias compared to the nominal value (red horizontal line) up to $dz=100 \ \mu$m for both the classical and the quantum case. The experimental points and their error bars agree with the confidence bandwidths provided by the simulations. However, the uncertainty on the estimated value is smaller for the quantum case. The quantum advantage, reported in Fig. \ref{Phase_estim}B,   is relatively constant in the range considered up to a 40$\%$. The estimated phase value drops down for higher defocusing distances because of the blurring of the image evident from the first row in Fig. \ref{pi}.
However, it is clear that in this region, the method does not provide useful reconstructions simply because the approximation of the derivative in Eq. (\ref{tie}) is no longer valid.

\begin{figure*}[th]
	\centering
	\includegraphics[height=0.40\textwidth, width=\textwidth]{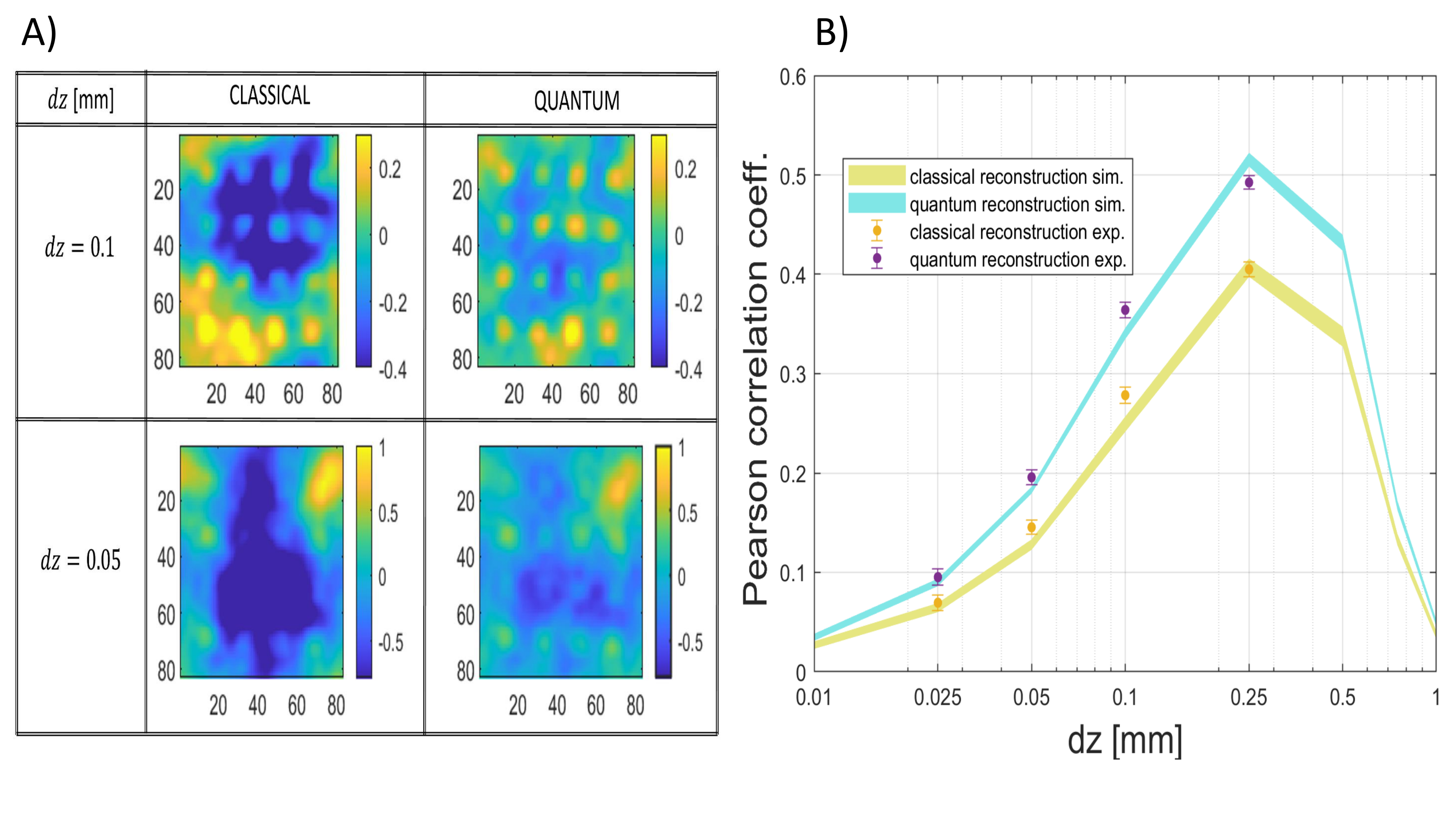}
	\caption{\textit{Single frame reconstruction of the squares pattern}. \textbf{A} Examples of classical and quantum reconstructions of the sample with squares in Fig.\ref{sample} (right-hand side) for two different defocussing distances. \textbf{B} Pearson correlation coefficient between the reconstructed phase image by a single intensity frame and the reference image as a function of the defocussing.}\label{Squares}
\end{figure*}

We have also tested a different object, the pattern of regular squares represented in Fig.  \ref{sample}. In Fig. \ref{Squares}A we report two examples of reconstructions, at $dz=50 \ \mu$m and $dz=100 \ \mu$m, respectively. In Fig. \ref{Squares}B, the Pearson coefficient is reported alongside the simulations. The quantum advantage is comparable to the one obtained for the $``\pi"$, showing its robustness and independence from the particular spatial shape of the sample. Although the quantitative analysis of the Pearson coefficient confirms a similar quantum advantage as the one reported in Fig. \ref{Pearson_vs_dz}, by looking at the images, it appears that the quantum advantage in the localization of dots could be even larger, indicating the possibility of significant advantages for specific tasks related to the recognition of finer spatial details.

In summary, these results demonstrate, for the first time, a significant advantage of quantum phase imaging, that can be further extended in the future with various potentially significant applications. 
 
\section{Conclusions}
 
Here, we have demonstrated a genuine quantum enhancement in non-interferometric quantitative phase imaging, showing that the spatially multi-mode quantum correlations can be used to reduce the detrimental effect of quantum noise in phase reconstruction. The present NIQPI scheme exports the classical methods known as the transport of intensity equation to the quantum regime, which provides real-time wide-field phase imaging and the quantitative local estimation of the phase. The last aspect is fundamental for many applications, providing reliable information on the object's internal parameters related to the phase. 
	
We point out that, compared to the imaging of an amplitude object \cite{Brida_2010,Samantaray_2017,RuoBerchera_2020,Losero_2018}, the propagation of the shot noise of the intensity measurement to the retrieved phase in the NIQPI is not as trivial. On the one side, the noise reduction allows reaching smaller defocussing distances for a better approximation of the derivative in the TIE, thus providing a more faithful reconstruction of the phase details.
On the other side, artifacts due to the noise appear at low spatial frequencies (see discussion in M$\&$M and Fig. \ref{pi}) and are known to affect mainly the reconstruction of slow phase curvature, which produces weaker signal intensity signals \cite{Paganin_2004}. In this work, in order to obtain a quantitative validation of the protocol, we studied binary phase objects with sharp borders. However, it is expected that for an object with smoother phase changes, e.g., biological samples, the quantum advantage can be even more significant.

\section{Materials and Methods}\label{MM}

 \subsection{Phase retrieval by TIE} 
 
 	A non-interferometric method \cite{Teague_1983} to retrieve the phase of an object consists of probing the object with a beam and measuring the intensity $I(\bm{x},z=0)$ at the object plane of coordinate $\bm{x}$ and its derivative along the propagation axis $z$. The derivative is computed by a finite difference of two measurements out-of-focus of a distance $d z$, $\frac{\partial}{\partial z}I(\bm{x},z)\approx\Delta I(\bm{x},d z)/(2 d z)$ with  $\Delta I(\bm{x},d z)= I(\bm{x},d z)-I(\bm{x},-d z)$. Under paraxial approximation, the phase is retrieved using the TIE reported in Eq. (\ref{tie}).

 	Using energy conservation considerations, this equation has been proven valid even with partially coherent sources \cite{Paganin_1998}. This feature makes the TIE approach perfectly suited for being used with light from SPDC, where transverse and longitudinal coherence lengths can be much smaller than the object size and the whole illuminating beam. This is not a secondary aspect since it is exactly due to the multimode nature of the emission that correlation shows a local character and shot noise can be removed pixel-by-pixel in the image. The solution of the Eq. (\ref{tie}) is unique provided that the on-focus intensity $I(\bm{x},0)$ and the intensity derivative along $z$ are known and the phase is continuous.\\ 
 	Following the analysis in \cite{Paganin_2004}, we assume that the intensity is varying sufficiently slowly that the effects of phase curvature dominate the intensity derivative, so that the right side of Eq. (\ref{tie}) can be safely approximated as $I_{0}\nabla^2\phi(\bm{x},0)$. Then, we consider for a moment that the only contribution to the finite difference $\Delta I(\bm{x},\delta z)$ is the noise fluctuation on the intensity measurement, $\sigma(\bm{x})$ . In this case, substituting the latter in Eq. (\ref{tie}), one has that the phase artifacts in the reconstruction due to the noise are:
  \begin{equation}
 -k\frac{\sigma(\bm{x})}{\sqrt{2} I_{0} \delta z}= \nabla_{\bm{x}}^2\phi_{noise}(\bm{x}).
 \label{noise artifact 1}
 \end{equation} 
 
 The noise is assumed independent in the two planes $+\delta z$ and $- \delta z$, so it has been combined in quadrature. The Eq. (\ref{noise artifact 1}) can be solved by taking the Fourier transform on both sides, leading to
 
\begin{equation}
k\frac{\tilde{\sigma}(\bm{q})}{4 \pi^{2}\sqrt{2} I_{0} \delta z |\bm{q}|^{2}}=\tilde{\phi}_{noise}(\bm{q})
\label{noise artifact 2}
\end{equation} 
where the tilde indicate the Fourier transform and $\bm{q}$ is the spatial frequency. The damping factor $|\bm{q}|^{2}$ of the higher frequencies at the denominator of Eq.   (\ref{noise artifact 2}) and the fact that the quantum noise (shot noise) has a flat white spectrum $\sigma_{SN} (\bm{q})= \sigma_{SN}$, indicate that the effect of shot noise is to generate artifacts especially at lower frequencies, which are not intrinsically suppressed by the phase retrieval algorithm. This noise at low-frequencies is evident in the single frame images reported in Fig. \ref{pi}. Moreover, in the direct propagation problem, higher frequencies of the phase object generate a stronger effect on the intensity. Thus, based on these remarks, the regions with rapid changes in the phase (higher frequency) are better reconstructed than the ones characterized by slow curvature.

\subsection*{Experimental details: Source, Sample, Detection}\label{Experimental details: source, sample, detection}	
\textit{Source}: In the experiment, we use SPDC in the low gain regime in which a photon of the pump beam (p) (CW laser @405nm), thanks to the interaction with a bulk beta-barium borate non-linear crystal as long as 15 mm, have a small probability of converting in a couple of photons, usually called signal (s) and idler (i), subject to conservation of energy, $\omega_p =  \omega_s + \omega_i$, and of momentum, $\textbf{k}_p =  \textbf{k}_s + \textbf{k}_i$. Thus, under the plane wave pump approximation, signal and idler photons are perfectly correlated in frequency and direction  $\bm{q}_s=-\bm{q}_i$ (assuming $\bm{q}_p=0$), although their individual spectrum is broadband both in space and frequency. In the far field, obtained at the focal plane of a thin lens in a $f-f$ configuration, a transverse mode $\bm{q}$ is mapped in a single transverse position $\bm{x}$ according to the transformation $(2 c f / \omega)\bm{q} \rightarrow \bm{x}$, so that momentum correlation translate in a position correlation, $\bm{x}_{s}=-\bm{x}_{i}$ (for degenerate frequency $\omega_{s}\approx\omega_{i}$). Signal and idler photons generate two symmetrical intensity noise patterns, and pairs of symmetric pixels of a camera will detect the same number of photons in the ideal lossless scenario in the same time window. Thus, quantum fluctuation affecting the object plane in the signal beam can be measured independently on the idler beam. 
The coherence time of the SPDC sources is typically of hundreds of fs and the spatial coherence in the far field is proportional to the inverse of the pump transverse size. The number of photons per spatial-temporal mode is very low, $\sim 10^{-8}$, in general, the time bandwidth of the detector is orders of magnitude smaller than the inverse of the coherence time. Although the single SPDC mode is thermal, in the limit above, the detected multi-thermal photon statistics are indistinguishable from a Poisson distribution \cite{Meda_2017}.

For a Gaussian distributed pump with angular full-width-half-maximum (FWHM) of $\Delta q$ the spatial cross-correlation is also Gaussian with FWHM of  $\Delta x=2\sqrt{2 \log 2} \sigma=(2 c f / \omega_p)\Delta q $: if a signal photon is detected in the position $\bm{x}_{s}$ the twin idler photon will be detected according to that Gaussian probability centered in $\bm{x}_{i}=-\bm{x}_{s}$.   In the experiment we have estimated $\Delta x\approx5\mu$m
 
\textit{Test sample}: The structures are etched on to a fused Silica glass window (WG41010-A, Thorlabs) with an anti-reflection coating on one side. The window is coated with positive PMMA resist and the design is exposed using electron beams. The exposed structures are developed using a MIBK-IPA solution. After development, the window is submerged in a buffered oxide etch for 30 seconds to etch the structures into the window. The etch depth is determined by the submergence time. The unexposed resist is then removed using acetone solution.

\textit{Detection}: We measure the SPDC emission and the effect of the phase object by imaging the far field of the source at the sensor of a CCD camera operated in the conventional linear mode. Each pixel delivers a count proportional to the number of incident photons. The proportionality coefficient is the product of the (electronic gain) that has been carefully calibrated and the quantum efficiency of the camera, nominally above 95\% @810 nm. The electronic readout noise is 4$e^{-}/(\text{pix}\cdot \text{frame})$. The number of photons detected per pixel per frame is $10^{3}$, where the integration time of the camera is set to $100$ ms, meaning that the shot noise dominates compared to the electronic noise.

Because of the finite cross-correlation area defined in the previous section of the M$\&$M, in order to collect most of the correlated photons, two symmetrically placed detectors (or pixels) must have areas larger than the cross-coherence area. Pixel size is 13 $\mu$m and a binning of $3 \times 3$ is performed to set the resolution to 5 $\mu$m at the object plane, which matches the measured cross-coherence area. Actually, the heralding efficiency $\eta$, i.e. the probability of detecting an idler photon conditioned to the prior detection of the twin signal photon,  depends on the pixel size $L$ and possible  misalignment $\bm{\Delta}$ of the two pixels compared to the optimal positions, according to this expression:

\begin{equation}
 \eta(L,\bm{\Delta})=\eta_0 L^{-2}\int_{L\times L} d\bm{x}_{s}\int_{L\times L} d\bm{x}_{i}\frac{1}{\sqrt{2\pi}\sigma}e^{-\frac{(\bm{x}_{i}+\bm{x}_{s}+\bm{\Delta})^{2}}{2\sigma^{2}}} 
\label{etac2}
\end{equation} 
where, $\eta_0$ is the single photon detection efficiency. As the the pixel size $L$ increases with respect to the coherence area $\Delta x$, we have that $\eta\longmapsto\eta_0$. 

\begin{figure}[th]
	\includegraphics[height=0.45\textwidth, width=0.5 \textwidth]{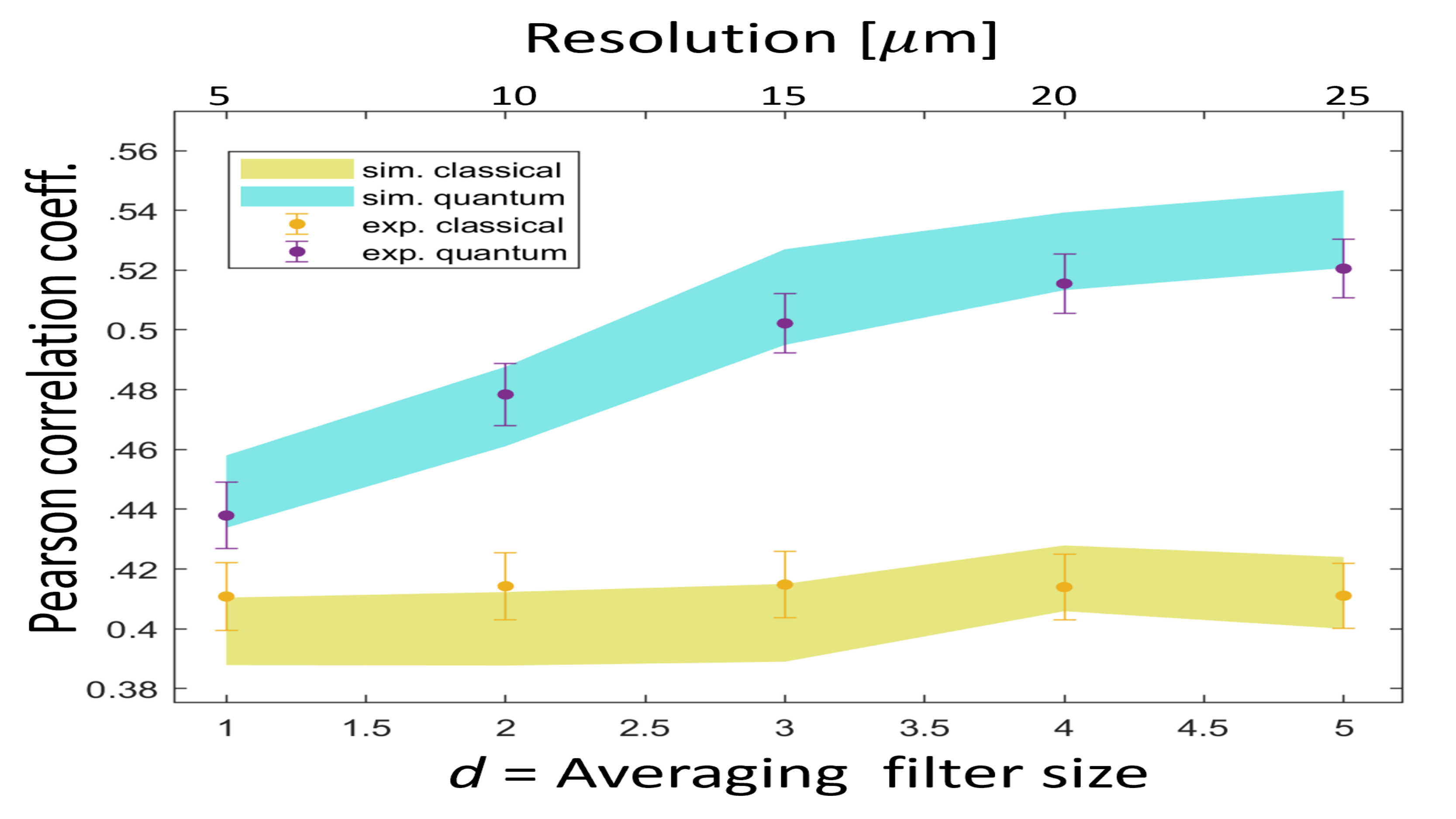} 
	\caption{\textit{Pearson correlation as a function of averaging filter size.} The purple (yellow) dots represent the values corresponding to the quantum (classical) experimental reconstructions. The quantum (classical) confidence bands at one standard deviation are also shown in 
			turquoise (yellow). }\label{Pearson_vs_d}
\end{figure}

As a consequence of that, in Eq. (\ref{noisered}), the noise reduction depends on the pixel size used for the measurement. This trade-off between the quantum advantage and the spatial resolution of the intensity measurement has been reported and analyzed in the context of sub-shot-noise imaging of amplitude objects \cite{Samantaray_2017, RuoBerchera_2020}. However, in the present imaging of pure phase objects, the resolution issue has less impact. In fact, as it is described in the first section of this M$\&$M, the solution of the TIE tends by itself to suppress the higher frequency component of the intensity perturbation. Thus, to some extent, a reduction of resolution in the intensity measurement does not affect the phase reconstruction. In the experiment, in order to increase the heralding efficiency, and thus the quantum enhancement, we use an averaging filter to the intensity image that substitutes the count in each pixel by the average of a neighborhood of size $d\times d \;\text{pix}^{2}$ around it. The quantum correlations are then enhanced because the effective integration area is larger, while the number of pixels in the final image is unvaried. In Fig. \ref{Pearson_vs_d}, we report the quality of the phase reconstruction, evaluated in terms of the Pearson correlation with the reference image in Fig. \ref{sample}, as a function of the averaging size. On the one hand, the quantum reconstruction is enhanced as expected when the effective resolution in the intensity measurement decreases ($d$ increases).
On the other hand, the classical reconstruction is unaffected, confirming that classically we do not have any negative issue related to the poorer resolution in the intensity pattern. In summary, moderate use of the averaging filter to enhance the quantum effects is perfectly legitimate in this context.

\subsection*{Model for the noise reduction} \label{Model for the noise reduction} 

According to the scheme in Fig. \ref{scheme}, the signal beam of SPDC probes the object, while the idler beam is used as a reference for the noise.
When the object is inserted with a defocusing distance $dz$, the photons in the signal beam are deflected, creating local depletion or accumulation of photons at the detection plane, and the perturbed intensity can be written as: 
\begin{equation}\label{I(z)}
	I_{s}(\bm{x},z)= I_{s}(\bm{x},0)-\Delta I_{-}(\bm{x}) +\Delta I_{+}(\bm{x}),
\end{equation}
where $I_{s}(\bm{x},0)$ is the unperturbed pattern and $\bm{x}$ indicates the position of a pixel. The quantity $\Delta I_{-}(\bm{x})$  $(\Delta I_{+}(\bm{x})$) represents the photons that are deflected out from (into) the position $\bm{x}$. 
From now on, to simplify the notation, the spatial average of the quantities is simply indicated by dropping the spatial dependence on $\bm{x}$. Since the total number of photons is conserved, the spatial average of the number of photons per pixel is unchanged, i.e. $I_{s}(z)=  I_{s}(0)$ and thus $ \Delta I_{-} =  \Delta I_{+}$. The loss of photons can be described as the action of a beam splitter of transmittance $1-\alpha$ (average value) so that, the quantum expectation value for the $\Delta I_{-}$ is simply $\langle\Delta I_{-}\rangle=\alpha\;\langle I_{s}(0) \rangle= \langle \Delta I_{+} \rangle$ \cite{Meda_2017}. In this work, we are interested in small perturbations that can be hidden or strongly affected by the quantum noise, so we will assume $\alpha\ll1$.

In order to reduce spatial intensity fluctuation we replace in the TIE the quantity in Eq. (\ref{I(z)}) with the one in Eq. (\ref{I(s-i)}) involving the idler measurement. 

The optimal factor $k_{opt}$ appearing there is chosen to minimize the residual fluctuation, by imposing $\frac{\partial}{\partial k} \langle \delta^2 I_{s-i}(\bm{x},z)\rangle=0$. We obtain,
\begin{eqnarray}\label{k_opt}
k_{opt}(\bm{x})&=&\frac{\langle \delta I_{s}(\bm{x}_s, z) \delta  I_{i}(\bm{x}_{i},0)\rangle}{\langle \delta^2  I_{i}(\bm{x}_{i},0)\rangle}, \\
\langle \delta^2 I_{s-i}(\bm{x}, z)\rangle&=& \langle \delta^2 I_{s}(\bm{x}_s, z) \rangle-\frac{\langle \delta I_{s}(\bm{x}_s, z) \delta I_{i}(\bm{x}_i, 0)\rangle^2}{\langle \delta^2 I_{i}(\bm{x}_i, 0)\rangle}.\nonumber
\end{eqnarray}

According to the Poisson distribution of the detected photon, we can replace the variance of the intensities appearing in Eq. (\ref{k_opt}) with the respective quantum mean values. In particular, by performing the spatial averaging, one gets  $\langle \delta^2  I_{i}(0)\rangle=\langle I_{i}(0)\rangle=\langle \delta^2  I_{s}(z)\rangle= \langle I_{i}(z)\rangle$. For the calculation of the covariance in Eq. (\ref{k_opt}), note that $I_{s}(\bm{x}_s, z)$ and $I_{i}(\bm{x}_i, 0)$ are correlated only for the fraction of photons that are not lost, namely not deviated from the path due to phase effect on the propagation along $z$. Thus, after spatial averaging \cite{Meda_2017}:

\begin{eqnarray}
	\langle\delta I_{s}(z) \delta I_{i}(0)\rangle&=& (1-\alpha) \langle\delta I_{s}(0) \delta I_{i}(0)\rangle\\ & =& \label{cov} \eta (1-\alpha)\langle I_{s}(0)\rangle.
\end{eqnarray}

The last equality is justified again using the Poisson hypothesis, and introducing the heralding efficiency $\eta$ that spoils the otherwise perfect signal-idler correlation. By using Eq. (\ref{cov}), and the Poisson hypothesis above, we can rewrite the Eq.s (\ref{k_opt}) as, 
\begin{eqnarray}\label{k_opt2}
k_{opt}&=&(1-\alpha) \,\eta\\
\langle \delta^2 I_{s-i}\rangle&=& \left[1-\left( 1-\alpha\right)^{2}\eta^{2} \right]\langle I_{s}(0)\rangle
\end{eqnarray}

\section{Acknowledgments}
This project 20FUN02 POLight has received funding from the EMPIR programme co-financed by the Participating States and from the European Union’s Horizon 2020 research and innovation programme. S.S. would like to acknowledge Dr. Iman E. Zadeh for his supervision for the sample fabrication.

\subsection*{Author contributions}
GO and IRB devised the scheme of NIQPI. PB participated in the realization of the setup and preliminary measurement with GO, while AP and CN performed the final experimental acquisitions, which were realised in the laboratories of the research sector directed by MG. The samples have been prepared and characterized by SS and SP. Simulations and data analysis have been carried out by GO and AP with the help of
IRB. IRB and MG supervised the whole project. IRB wrote the paper with the contribution of all authors.

\subsection*{Competing Interests}
The authors declare no competing interest.
\subsection*{Data availability} All data needed to evaluate the conclusions are reported in the paper. Further data, for reproducibility of the results, will be available in a public repository linked to the published paper.

\bibliography{bib}	

\end{document}